\documentstyle[11pt,a4]{article}
{

\hoffset=15mm
\textwidth=140mm
\clubpenalty 10000
\widowpenalty 10000

\author{Afraimovich~E.~L., Perevalova~N.~P.,
        Plotnikov~A.~V., Uralov~A.~M. \\
        Institute of Solar-Terrestrial Physics SD RAS,\\
        p.~o.~box~4026, Irkutsk, 664033, Russia\\
        fax: +7 3952 462557; e-mail:~afra@iszf.irk.ru}

\title{The shock-acoustic waves generated by earthquakes }

\date{}

\begin{document}
\sloppy
\maketitle
\begin{abstract}
We investigate the form and dynamics of shock-acoustic waves
generated by earthquakes. We use the method for detecting and
locating the sources of ionospheric impulsive disturbances, based on
using data from a global network of receivers of the GPS navigation
system and requiring no a priori information about the place and time of
associated effects. The practical implementation of the method is
illustrated by a case study of earthquake effects in Turkey (August 17,
and November 12, 1999), in Southern Sumatera (June 4, 2000), and off
the coast of Central America (January 13, 2001). It was found that in
all instances the time period of the ionospheric response
is 180-390 s, and the amplitude exceeds by a
factor of two as a minimum the standard deviation of background
fluctuations in total electron content in this range of periods under quiet
and moderate geomagnetic conditions. The elevation of the wave
vector varies through a range of 20-44${}^\circ$, and the phase
velocity (1100-1300 m/s) approaches the sound velocity at the heights
of the ionospheric F-region maximum. The calculated (by neglecting
refraction corrections) location of the source roughly corresponds to
the earthquake epicenter. Our data are consistent with the present
views that shock-acoustic waves are caused by a piston-like movement
of the Earth surface in the zone of an earthquake epicenter.
\end{abstract}

\section{Introduction}
\label{EARTHQ-sect-1}

A plethora of publications have been devoted to the study of the
ionospheric response to disturbances arising from impulsive forcing on
the Earth's atmosphere. It was found that in many cases a high
proportion of the energy of the initial atmospheric disturbance is
concentrated in the acoustic shock wave. Large earthquakes also
provide a natural source of impulsive forcing.

These investigations have also important practical implications
since they furnish a means of substantiating reliable signal
indications of technogenic effects (among them, unauthorized),
which is necessary for the construction of an effective global
radiophysical system for detection and localization of these
effects. Essentially, existing global systems of such a purpose
use different processing techniques for infrasound and seismic
signals. However, in connection with the expansion of the
geography and of the types of technogenic impact on the
environment, very challenging problems heretofore have been those
to improve the sensitivity of detection and the reliability of
measured parameters of the sources of impacts, based also on
using independent measurements of the entire spectrum of signals
generated during such effects.

To solve the above problems requires reliable information about
fundamental parameters of the ionospheric response to the shock
wave, such as the amplitude and the form, the period, the phase
and group velocity of the wavetrain, as well as angular
characteristics of the wave vector. Note that for naming
the ionospheric response of the shock wave, the literature uses
terminology incorporating a different physical interpretation,
among them the term `shock-acoustic wave' (SAW)~(Nagorsky, 1998).
For convenience, in this paper we shall use this term
despite the fact that it does not reflect essentially the
physical nature of the phenomenon.

There is a wide scatter of published data on the main parameters of
SAWs generated during industrial explosions and earthquakes. The
oscillation period of the ionospheric response of SAWs varied from 30
to 300 sec., and the propagation velocity fluctuated from 700 to 1200
m/s (Nagorsky, 1998; Fitzgerald, 1997; Calais et al., 1998;
Afraimovich et al., 1984; Blanc and Jacobson, 1989).

The lack of comprehensive, reliable data on SAW parameters is due
primarily to the limitations of existing experimental methods and
detection facilities. The main body of data was obtained by measuring
the frequency Doppler shift at vertical and oblique-incidence
ionospheric soundings in the HF range (Nagorsky, 1998; Afraimovich
et al., 1984; Jacobson and Carlos, 1994). In some instances the
sensitivity of this method is sufficient to detect the SAW reliably;
however, difficulties emerge for localizing the region where the
detected signal is generated. These problems are caused by the
multiple-hop character of HF signal propagation. This gives no way of
deriving reliable information about the phase and group velocities of
the SAW propagation, estimating angular characteristics of the wave
vector, and, further still, of localizing the SAW source.

Using the method of transionospheric sounding with VHF radio
signals from geostationary satellites, a number of experimental
data on SAW parameters were obtained in measurements of the
Faraday rotation of the plane of signal polarization which is
proportional to a total electron content (TEC) along the line
connecting the satellite-borne transmitter with the receiver
(Li et al., 1994).

A common drawback of the above-mentioned methods when
determining the SAW phase velocity is the necessity of knowing
the time of events since this velocity is inferred
from the SAW delay with respect to the time of events assuming that
the velocity is constant along the propagation path, which is
quite contrary to fact.

For determining the above-mentioned rather complete set of SAW
parameters, it is necessary to have appropriate spatial and temporal
resolution which cannot be ensured by existing, very sparse, networks
of ionosondes, oblique-incidence radio sounding paths, and incoherent
scatter radars.

The advent of the Global Positioning System (GPS) plus the subsequent
creation of extensive networks of GPS stations (at least 757 sites as of
November 2000), with their data being now available via the Internet,
opened up a new era in remote ionospheric sensing.

Currently some authors have embarked on an intense development of
methods for detecting the ionospheric response of strong
earthquakes (Calais and Minster, 1995), rocket launchings (Calais
and Minster, 1996), and industrial surface explosions
(Fitzgerald, 1997; Calais et al., 1998).
In the cited references the SAW phase velocity was determined by
the `crossing' method by estimating the time delay of SAW arrival
at subionospheric points corresponding to different GPS
satellites observed at a given time. However, the accuracy of
such a method is rather low because the altitude at which the
subionospheric points are specified, is determined in a crude
way.

The goal of this paper is to describe a method for determining
parameters of the SAW generated by earthquakes (including the phase
velocity, angular characteristics of the SAW wave vector, the direction
towards the source, and the source location) using GPS-arrays whose
elements can be chosen out of a large set of GPS stations from the
global GPS network.

Section~2 presents a description of the
experimental geometry, and general information about the earthquakes
under consideration. The proposed method is briefly outlined in
Section~3. Results of measurements of SAW parameters from different GPS
arrays during earthquakes are presented in Section~4. Section~5 is devoted
to the discussion of experimental results, including analytical
simulation results.

\section{The geometry and general characterization of experiments}
\label{EARTHQ-sect-2}

Detection results on two earthquakes in Turkey (August 17, and
November 12, 1999), in Southern Sumatera (June 4, 2000), and off the
coast of Central America (January 13, 2001) are presented below. The
information about the earthquakes, was acquired via the Internet
`http://earthquake.usgs.gov'. General information about these
earthquakes is presented in Table~1 (including
the time of the main shock $t_{0}$ in the universal time UT, the
position of the earthquake epicenter, depth, the magnitude, as well as
the level of geomagnetic disturbance from the data on
$Dst$-variations). It was found that the deviation of $Dst$ for the
selected days was quite moderate, which enabled the SAWs to be
identified.

Fig.~1 illustrates the experimental geometry
during the earthquakes in Turkey -- a), off the coast of Central
America -- b), and in Southern Sumatera -- c).

In spite of the small number of GPS stations in the earthquake area, we
were able to use a sufficient number of them for the implementation of
the proposed method.

Table~2 presents the geographic coordinates of
the GPS stations used as GPS array elements.

\section{Methods of determining shock-acoustic wave characteristics
using GPS-arrays}
\label{EARTHQ-sect-3}

The standard GPS technology provides a means for wave disturbances
detecion based on phase measurements of TEC at each of spaced
two-frequency GPS receivers. A method of reconstructing TEC
variations was detailed and validated in a series of publications (Calais
and Minster, 1995, 1996; Fitzgerald, 1997). We reproduce here only
the final formula for the total electron content (I)

\begin{equation}
\label{EARTHQ-eq-1}
I=\frac{1}{40{.}308}\frac{f^2_1f^2_2}{f^2_1-f^2_2}
                           [(L_1\lambda_1-L_2\lambda_2)+const+nL]
\end{equation}
where $L_1\lambda_1$ and $L_2\lambda_2$ are additional paths of
the radio signal caused by the phase delay in the
ionosphere,~(m); $L_1$ and $L_2$ represent the number of phase
rotations at the frequencies $f_1$ and $f_2$; $\lambda_1$ and
$\lambda_2$ stand for the corresponding wavelengths,~(m);
$const$ is the unknown initial phase ambiguity,~(m); and $nL$~
are errors in determining the phase path,~(m).

Phase measurements in the GPS can be made with a high degree of
accuracy corresponding to the error of TEC determination of at least
$10^{14}$~m${}^{-2}$ when averaged on a 30-second interval, with
some uncertainty of the initial value of TEC, however. This makes
possible detection of ionization irregularities and wave processes in the
ionosphere over a wide range of amplitudes (up to $10^{-4}$ of the
diurnal TEC variation) and periods (from 24 hours to 5 min). The unit
of TEC, $TECU$, is equal to $10^{16}$~m${}^{-2}$, it is
commonly accepted in the literature, and will be used in the following.

A convenient way of determining the ionospheric response delay of the
shock wave involves the frequency Doppler shift $F$ from TEC series
obtained by formula~(\ref{EARTHQ-eq-1}). Such an approach is also
useful in comparing TEC response characteristics from the GPS data
with those obtained by analyzing VHF signals from geostationary
satellites, as well as in detecting the shock wave in the HF range. To
an approximation sufficient for the purpose of our investigation, a
corresponding relationship was obtained by K. Davies (1969)

\begin{equation}
\label{EARTHQ-eq-2}
F=13{.}5\cdot  10^{-8}I'_t/f
\end{equation}
where $I'_t$ stands for the time derivative of TEC. Relevant results
derived from analyzing the $F(t)$-variations calculated for the
`reduced' frequency of 136~MHz are discussed in
Section~\ref{EARTHQ-sect-4}.

The correspondence of space-time phase characteristics, obtained
through transionospheric soundings, with local characteristics of
disturbances in the ionosphere was considered in detail in a wide
variety of publications (Afraimovich et al., 1992; Mercier and
Jacobson, 1997) and is not analyzed at length in this study. The most
important conclusion of the cited references is the fact that, as for the
extensively exploited model of a `plane phase screen' disturbances
$\Delta I(x,y,t)$ of TEC faithfully copy the horizontal part of the
corresponding disturbance of local electron concentration $\Delta
N(x,y,z,t)$, independently of the angular position of the source, and
can be used in experiments for measuring the TEC disturbances.

However, the TEC response amplitude experiences a strong azimuthal
dependence caused by the integral character of a transionospheric
sounding. As a first approximation, the transionospheric sounding
method is responsive only to Traveling Ionospheric Disturbances
(TIDs) with the wave vector $\boldmath K_t$ perpendicular to the
direction $\boldmath r$ wich is along the Line-of-Sight (LOS) from
the receiver to the satellite. A corresponding condition for elevation
$\theta$ and azimuth $\alpha$ of an arbitrary wave vector $\boldmath
K_t$ normal to the direction $\boldmath r$, has the form

\begin{equation}
\label{EARTHQ-eq-3}
\theta=\arctan(-\cos(\alpha_s-\alpha)/\tan\theta_s)
\end{equation}
where $\alpha_s$ is the azimuthal angle measured east to north,
and $\theta_s$ is the angle of elevation of the satellite at the receiver.

We used formula~(\ref{EARTHQ-eq-3}) to determine the elevation
$\theta$ of $\boldmath K_t$ from the known mean value of azimuth
$\alpha$ by Afraimovich et al. (1998) -- see Sections
~\ref{EARTHQ-sect-3.2} and~\ref{EARTHQ-sect-4}.

\subsection{Detection and determination of the horizontal phase
velocity and the direction of the SAW phase front along the ground by
GPS-arrays}
\label{EARTHQ-sect-3.1}

In the simplest form, space-time variations of the TEC
$\Delta I(t,x,y)$ in the ionosphere, at each given time $t$ can be
represented in terms of the phase interference pattern that moves
without a change in its shape (the solitary, plane travelling
wave)

\begin{equation}
\label{EARTHQ-eq-4}
\Delta I(t,x,y)=\delta\sin(\Omega t-K_xx-K_yy+\varphi_0)
\end{equation}
where $\delta$, $K_x$, $K_y$, $\Omega$, are the amplitude,
the x- and y-projections of the wave vector {\boldmath $K$}, and
the angular frequency of the disturbance, respectively;
$T=2\pi/\Omega$, $\Lambda=2\pi/|\boldmath K|$ is its period and wavelength;
and $\varphi_0$~ is the initial phase of the disturbance. The
vector $\boldmath K$ is a horizontal projection of the
full vector $\boldmath K_t$.

At this point, it is assumed that in the case of small spatial
and temporal increments (the distances between GPS-array sites
are less than the typical spatial scale $\Lambda$ of TEC variation, and the
time interval between counts is less than the corresponding time
scale), the influence of second derivatives can be neglected. The
following choices of GPS-arrays all meet these requirements.

We now summarize briefly the sequence of data processing
procedures. Out of a large number of GPS stations, three sites
(A, B, C) are selected, within distances not exceeding
about one-half of the expected wavelength $\Lambda$ of the
perturbation. Site B is taken to be the center of a topocentric
reference frame whose axis $x$ is directed east, and the axis
$y$ is directed north. The receivers in this frame of
reference have the coordinates €($x_A$, $y_A$), '(0,0), '($x_C$,
$y_C$). Such a configuration of the GPS receivers
represents the GPS-array with a minimum number of the required
elements. In regions with a dense network of GPS sites, we can
obtain a large variety of GPS-arrays of a different configuration
enabling the acquired data to be checked for reliability; in this
paper we have exploited this possibility.

The input data include series of slant TEC values $I_A(t)$, $I_B(t)$,
$I_'(t)$, as well as corresponding series of elevation values
$\theta_s(t)$ and the azimuth $\alpha_s(t)$ of the LOS.
For determining SAW characteristics, continuous series of
measurements of $I_A(t)$, $I_B(t)$, $I_C(t)$ are selected with a
length of at least a one-hour interval which includes the time of a
earthquake.

To eliminate spatio-temporal variations of the regular ionosphere, as
well as trends introduced by orbital motion of the satellite, a procedure
is used to remove the trend involving a preliminary smoothing of the
initial series with the selected time window. This procedure is better
suited to the detection of a single pulse signal ($N$-wave) than the
frequently used band-pass filter (Li et al., 1994; Calais and Minster,
1995, 1996; Fitzgerald, 1997; Calais et al., 1998). A limitation of the
band-pass filter is the oscillatory character of the
response which prevets from reconstructing the form of the $N$-wave.

Elevation $\theta_s(t)$ and azimuth $\alpha_s(t)$ values of the LOS are used
to determine the location of the subionospheric point, as well as to
calculate the elevation $\theta$ of the wave vector $\boldmath K_t$ of
the disturbance from the known azimuth $\alpha$ (see
formula~(\ref{EARTHQ-eq-3})).

The most reliable results from the determination of SAW
parameters correspond to high values of elevations $\theta_s(t)$
of the LOS because sphericity effects become
reasonably small. In addition, there is no need to convert
the slant TEC $\Delta I(t)$ to a `vertical' value.
In this paper, all results were obtained for elevations
$\theta_s(t)$ larger than 30{}$^\circ$.

Since the distance between GPS-array elements (from several tens
of kilometers to a few hundred kilometers) is much smaller than
that to the GPS satellite (over 20000~km), the array geometry at
the height of the ionosphere is identical to that on the ground.

Fig.~2a shows typical time dependencies of an
slant TEC $I(t)$ at the GPS-array BSHM station
near the area of the earthquake of August 17, 1999. (heavy
curve), one day before and after the earthquake (thin lines). For the
same days; panel b shows TEC variations $\Delta I(t)$
after removal of a linear trend and  smoothing by averaging over a
sliding window of  5-min.
Variations in frequency Doppler shift $F(t)$, `reduced' to the sounding
signal frequency of 136~MHz, for three sites of the array (KATZ
BSHM GILB) on August 17, 1999, are presented in panel c.

Fig.~2 shows that fast $N$-shaped
oscillations, with a typical period of about 390~s, are
distinguished among slow TEC variations. The oscillation amplitude
(up to 0{.}12~$TECU$) is far in excess of the background TEC
fluctuation intensity as seen on the days before and after the
earthquake. Variations in frequency Doppler shift $F(t)$ for spatially
separated sites (KATZ BSHM GILB) are well correlated but are shifted
relative to each other by an amount well below the period, which
permits the SAW propagation velocity to be unambiguously
determined. The 30~s sampling rate of the GPS data is not quite
sufficient for determining small shifts of such signals with an adequate
accuracy for different sites of the array. Therefore, we used a parabolic
approximation of the $F(t)$-oscillations in the neighborhood of
minimum $F(t)$, which is quite acceptable when the signal/noise ratio
is high.

Taking into account the good signal/noise ratio (better than~1), and
knowing the coordinates of the array sites A,~B and~C, we determine
the horizontal projection of the phase velocity $V_h$ from time shifts
$t_p$ of a maximum deviation of the frequency Doppler shift $F(t)$.
Preliminarily measured shifts are subjected to a
linear transformation with the purpose of calculating shifts for
sites spaced relative to the central site northward $N$ and
eastward $E$. This is followed by a calculation of the $E$- and
$N$-components of $V_x$ and $V_y$, as well as the direction
$\alpha$ in the range of angles 0--360${}^\circ$ and
the modulus $V_h$ of the horizontal component of the SAW phase
velocity

\begin{equation}
\label{EARTHQ-eq-5}
\begin{array}{rl}
\alpha&=\arctan(V_y/V_x)\\ V_h&=|V_xV_y|(V_x^2+V_y^2)^{-1/2}
\end{array}
\end{equation}
where $V_y$, $V_x$ are the velocities with which the phase front
crosses the axes $x$ and $y$. The orientation $\alpha$ of the
wave vector $\boldmath K$, which is coincident with the
propagation azimuth of the SAW phase front, is calculated
unambiguously in the range 0--360${}^\circ$ subject to
the condition that $arctan(V_y/V_x)$ is calculated having regard
to the sign of the numerator and denominator.

The above method for determining the SAW phase velocity neglects
the correction for orbital motion of the satellite because the estimates
of $V_h$ obtained below exceed an order of magnitude as a minimum
the velocity of the subionospheric point at the height of the ionosphere
for elevations $\theta_s>30^\circ$~ (Afraimovich et al., 1998).

Obviously the method presented above can be used if the distance
between GPS stations is much shorter than the TEC disturbance
wavelength $\Lambda$ and the distance from the earthquake
epicenter to the array. This corresponds to the detection
condition in the far-field zone.

From the delay $\Delta t=t_p-t_0$ and the known path length
between the earthquake focus and the subionospheric point we
calculated also the SAW mean velocity $V_a$ in order to compare
our estimates of the SAW phase velocity with the usually used
method of measuring this quantity.

\subsection{Determination of the wave vector elevation $\theta$
and the velocity modulus $V_t$ of the shock wave}
\label{EARTHQ-sect-3.2}

Afraimovich et al. (1992) showed that for the Gaussian ionization
distribution the TEC disturbance amplitude ($M$) is determined by
the aspect angle $\gamma$ between the vectors $\boldmath K_t$ and
$\boldmath r$, as well as by the ratio of the wavelength of the
disturbance $\Lambda$ to the half-thickness of the ionization
maximum $h_d$

\begin{equation}
\label{EARTHQ-eq-6}
M\propto\exp\left(-\frac{\pi^2h_d^2\cos^2\gamma}
                        {\Lambda^2\cos^2\theta_s}\right).
\end{equation}

In the case under consideration (see below), for a phase
velocity on the order of 1~km/s and for a period of about 200~s, the
wavelength $\Lambda$ is comparable with the half-thickness of the
ionization maximum $h_d$. When the elevations $\theta_s$ are
30{}$^\circ$, 45{}$^\circ$, 60{}$^\circ$ , the `beam-width' $M(\gamma)$
at 0{.}5 level is, respectively, 25{}$^\circ$, 22{}$^\circ$ and
15{}$^\circ$. If $h_d$ is twice as large as the wavelength, then
the beam tapers to 14{}$^\circ$, 10{}$^\circ$ and 8{}$^\circ$,
respectively.

The beam-width is sufficiently small that the aspect
condition~(\ref{EARTHQ-eq-3}) restricts the number of beam
trajectories to the satellite, for which it is possible to detect
reliably the SAW response in the presence of noise (near the
angles $\gamma=90^\circ$ ). On the other hand,
formula ~(\ref{EARTHQ-eq-3}) can be used to determine the
elevation $\theta$ of the wave vector $\boldmath K_t$ of the
shock wave at the known value of the azimuth $\alpha$
(Afraimovich et al., 1998). Hence the phase velocity modulus
$V_t$ can be defined as

\begin{equation}
\label{EARTHQ-eq 7}
V_t=V_h\cdot cos(\theta)
\end{equation}

The above values of the width $M(\gamma)$ determine the error of
calculation of the elevations $\theta$ (of order 20{}$^\circ$ to the
above conditions).

\subsection{Determining the position of the SAW source without
regard for refraction corrections}
\label{EARTHQ-sect-3.3}

The ionospheric region that is responsible for the main
contribution to TEC variations lies in the neighborhood of the
maximum of the ionospheric $F$-region, which does determine the
height $h_{i}$ of the penetration point. When selecting
$h_{i}$, it should be taken into consideration that the
decrease in electron density with height above the main maximum
of the $F_2$-layer proceeds much slower than
below this maximum. Since the density distribution with height is
essentially a `weight function' of the TEC response to a wave
disturbance (Afraimovich et al., 1992), it is appropriate to use,
as $h_{i}$, the value exceeding the true height of the layer
$h_{m}F2$ maximum by about 100~km. $h_{m}F2$ varies
between 250 and 350~km depending on the time of day
and on some geophysical factors which, when necessary, can be
taken into account if additional experimental data
and current ionospheric models are available. In all calculations that
follow, $h_{i}=400$~km is used.

To a first approximation, it can be assumed that the imaginary detector,
which records the ionospheric SAW response in TEC variations is
located at this altitude. The horizontal extent of the detection region,
which can be inferred from the propagation velocity of the
subionospheric point as a consequence of the orbital motion of the GPS
satellite (on the order of 70~--150~m/s; see Pi at al., 1997), and from the
SAW period (on the order of 200~s~--- see
Section~\ref{EARTHQ-sect-4}), does not exceed 20--40~km, which
is far smaller than it `vertical size' (on the order of the half-thickness of
the ionization maximum $h_d$).

From the GPS data we can determine the coordinates $X_s$ and
$Y_s$ of the subionospheric point in the horizontal plane $X0Y$
of a topocentric frame of reference centered on the point B(0,0)
at the time of a maximum TEC deviation caused by the arrival of
the SAW at this point. Since we
know the angular coordinates $\theta$ and $\alpha$ of the wave
vector $\boldmath K_t$, it is possible to determine the location
of the point at which this vector intersects the horizontal plane
$X'0Y'$ at the height $h_w$ of the assumed source. Assuming a
rectilinear propagation of the SAW from the source to the
subionospheric point and neglecting the Earth sphericity, the coordinates
$X_w$ and $Y_w$ of the source in a topocentric frame of reference
can be defined as

\begin{equation}
\label{EARTHQ-eq-8}
X_w=X_p-\left(h_{i}-h_w\right)
             \frac{\cos\theta\sin\alpha} {\sin\theta}
\end{equation}
\begin{equation}
\label{EARTHQ-eq-9}
Y_w=Y_p-\left(h_{i}-h_w\right)
             \frac{\cos\theta\cos\alpha} {\sin\theta}
\end{equation}

The coordinates $X_w$ and $Y_w$, thus obtained, can readily be
recalculated to the values of the latitude and longitude
($\phi_w$ and $\lambda_w$) of the source.

For SAW generated during earthquakes, industrial explosions and
underground tests of nuclear devices, $h_w$ is taken to be equal
to 0 (the source lying at the ground level).

\section{Results of measurements}
\label{EARTHQ-sect-4}

Hence, using the transformations described in
Section~\ref{EARTHQ-sect-3}, we obtain the parameters set
determined from TEC variations and characterizing the SAW
(see Table~3).

Let us consider the results derived from analyzing the
ionospheric effect of SAW during earthquake August 17, 1999
obtained at the array (KATZ, BSHM, GILB) for PRN 6 (at the left
of Fig.~2, and line~2 in Table~3).

In this case the delay of the SAW response with respect to the
time of the earthquake is 20~min (DAY 229). The SAW has the form
of an $N$-wave with a period $T$ of about 390~s and an amplitude
$A_I=0.12~TECU$, which is an order of magnitude larger than TEC
fluctuations for background days (DAY 228, DAY 230). It should be
noted, however, that this time interval was characterized by a
very low level of geomagnetic activity (-14~nT).

The amplitude of a maximum frequency Doppler shift $A_F$ at the
`reduced' frequency of 136~MHz was found to be 0.04~Hz. In view
of the fact that the shift $F$ is inversely proportional to the
sounding frequency squared (Davies, 1969), this corresponds to a
Doppler shift at the working frequency of 13.6~MHz and the
equivalent oblique-incidence sounding path of about $A_F=4$~Hz.

Solid curves in Fig.~1a represent the trajectories of
the subionospheric points for each GPS satellite at the height
$h_{i}$=400 km during the time interval 0.0--0.8 UT for August
17, 1999, and 17.0--17.4 UT for November 12, 1999.
Dark diamonds along the trajectories correspond
to the coordinates of the subionospheric points at time $t_p$ of a
maximum deviation of the TEC (Fig.~2b and
Fig.~2e). Crosses show the positions of the
earthquake epicenters. Asterisks mark the source location at 0 km
altitude inferred from the data from the GPS arrays. Numbers at the
asterisks correspond to the respective day numbers. Straight dashed
lines that connect the expected source and the subionospheric point
represent the horizontal projection of the wave vector
$\boldmath K_t$.

The azimuth $\alpha$ and elevation $\theta$ of the wave vector
$\boldmath K_t$ whose horizontal projection is shown in
Fig.~1a by a dashed line and is marked by
$\boldmath K_1$, are 154{}$^\circ$ and 26{}$^\circ$,
respectively. The horizontal component and the modulus of the
phase velocity were found to be $V_h=1307$~m/s and
$V_t=1174$~m/s. The source coordinates at 0~km altitude were
determined as $\phi_w=39.1^\circ$ and $\lambda_w=25.9^\circ$. The
áalculated (by neglecting refraction corrections) location of the
source roughly was close the earthquake epicenter.

The `mean' velocity of about $V_a=870$~m/s, determined in a usual
manner from the response delay with respect to the start, was smaller
then the phase velocity $V_t$.
Conceivably this is associated with an added delay of the response as a
consequence of the refraction distortions of the SAW path along the
LOS which were neglected in this study.

Similar results for the array (KABR, TELA, GILB) and PRN 30 were
also obtained for the earthquake of November 12, 1999. They
correspond to the projection of the vector $\boldmath K_2$ in
Fig.~1a, the time dependencies in Fig.~2 at the right, and line 7 in
Table~3. The only point deserving mention here
is that the SAW amplitude was somewhat smaller than that for the
earthquake of August 17, 1999. With an increased level of
geomagnetic activity (~-44 nT), this led to a smaller (compared
with August 17, 1999) signal/noise ratio; yet this did not
preclude reliable estimates of the SAW parameters.

A comparison of the data for both earthquakes showed a reasonably
close agreement of SAW parameters irrespective of the level of
geomagnetic disturbance, the season, and the local time.

To convince ourselves that the determination of the main parameters of
the SAW form and dynamics is reliable for the earthquakes analyzed
here, in the area of the earthquake we selected different
combinations of three sites out of the sets of GPS stations available to
us, and these data were processed with the same processing
parameters. Relevant results (including the average results for the sets
$\Sigma$), presented in Table~3 and in
Fig.~1a (SAW source position), show that the
values of SAW parameters are similar, which indicates a good stability
of the data, irrespective of the GPS-array configuration.

The aspect condition~(\ref{EARTHQ-eq-3}), corresponding to a
maximum amplitude of the TEC response to the transmission of SAWs,
was satisfied quite well for this geometry, simultaneously for all
stations. This is confirmed by a high degree of correlation of the SAW
responses at the array elements (Fig.~2c and
Fig.~2f), which made it possible to obtain different
sets of triangles out of the six GPS stations available to us.

The relative position of the GPS stations was highly convenient for
determining the SAW parameters during the earthquakes in Turkey,
and met the implementation conditions for the method described in
Section~\ref{EARTHQ-sect-3.1}. Thus, the distance between stations
(100-150 km at most) did not exceed the SAW wavelength of about
200-300 km, and was far less than the distance from the epicenter to
the array (1000 km).

Let us consider the results derived from analyzing the
ionospheric effect of SAW during earthquake January 13, 2001,
obtained at the array (TEGU, MANA, ESTI) for PRN13 (at the left
of Fig.~3 and line 11 in Table~3).

In this case the delay of the SAW response with respect to the
time of the earthquake is 15~min (DAY 013). The SAW has the form
of oscillations with a period $T$ of about 270~s and an
amplitude $A_I=0.2~TECU$, which is an order of magnitude larger
than TEC fluctuations for background days (DAY 012, DAY 014). It
should be noted, that this time interval was characterized by a
very low level of geomagnetic activity (4~nT).

Similar results for the array (SAMP, NTUS, BAKO) and PRN 03
were also obtained for the  earthquake  of  June 4, 2000.
(at the right of Fig.~3 and line 10 in
Table~3).

Note that in this case the response amplitude exceeded twice, as a
minimum, that for the other events under consideration. It is not
improbable that this is due to the maximum magnitude of the
earthquake in Southern Sumatera (see Table~1).

Unfortunately, because of the inadequately well-developed network of
stations, for the earthquakes in Southern Sumatera (June 4, 2000) and
off the coast of Central America (January 13, 2001) for each of these
events it was impossible to select arrays meeting the applicability
conditions of the method for determining the SAW wave vector
parameters described in Section~\ref{EARTHQ-sect-3.1}.

It is evident from the geometry in Fig.1b and
Fig.~1c that the earthquake epicenters lay inside
the GPS arrays, which does not meet the far-field zone condition. It is
possible that this is also responsible for the form of the response
(presence of strong oscillations) which differs from the N-form of the
response for the earthquakes in Turkey. On this basis, for these
earthquakes we can point out only the very fact of reliable detection of
the response, and determine its amplitude $A_I$, typical period $T$,
delay $\Delta t=t_p-t_0$, and velocity $V_a$ (see
Table~3). The values of these quantities were
close to the data obtained for the earthquakes in Turkey.

\section{Discussion}
\label{EARTHQ-sect-5}

The data of Table~3 are quite sufficient to
estimate the position of the disturbances under discussion in the
diagnostic diagram of the atmospheric waves. Specifically, the
values of the characteristic periods of bipolar signals,
presented in Fig.~2b and Fig.~2e, are $T_1 \approx$ 300~s, and $T_2
\approx$ 200~s. The wave vectors are at the angles
$\beta_1 \approx 70^\circ$ and $\beta_2 \approx 50^\circ$
with vertical.
At the height $z \approx$ 400~km, in turn, the periods,
corresponding to local frequencies of the acoustic cut-off
$\omega_a$ and the Brunt-V\"ais\"al\"a $\omega_b$, are:
$T_a=2\pi/\omega_a \approx$ 950~s, $T_b=2\pi/\omega_b \approx$
1050~s.

In an isothermal atmosphere, only harmonics with periods larger than
$T_b/sin~\beta$ can be assigned to the branch of internal gravity
waves. This value exceeds significantly the values of $T_{1,2}$.
Furthermore, $T_{1,2}$ is considerably smaller than $T_a$. Hence we
can contend that under conditions of the real (nonisothermal)
atmosphere the disturbances under discussion almost entirely pertain to
the branch of acoustic waves.

Since in (short period) acoustic waves the variation $\Delta N$
in neutral density
$N$ satisfies the relation $\Delta N /N \approx u / C$
($u$ is the gas velocity in the wave,  and
$C$  is  the local velocity of sound),  it is possible to make a lower
estimate  of  the  intensity $u/C$ of   the   waves at   the   expected
altitudes $z_{eff} \approx 350$--400~km:~~
$\Delta N  /  N  \approx  \Delta  N_e  /  N_e  \approx \Delta I / I \approx
0.02-0.04$.
In this estimation,  the influence
of  the  magnetic  field  is  omitted,  and  the  index  $e$   refers   to
electrons.  The  actual  intensity  of  the  acoustic  wave  must  be
higher than the estimated value presented above.

In the case of an earthquake, the movements of the terrestrial surface
are plausible sources of acoustic waves. A generally known source of
the first type is the Rayleigh surface wave propagating from the
epicentral zone. The ground motion in the epicentral area is the
source of the second type. Let us discuss these possibilities.

\subsection{The Rayleigh wave}
\label{EARTHQ-sect-5.1}

The phase  propagation  velocity  of  the Rayleigh wave
$V_R \approx 3.3$~km/s.
Since $V_R \gg C_0$, where $C_0 \approx 0.34$~km/s is
the sound velocity
at the ground, only acoustic waves can be emitted (Golitsyn and
Klyatskin, 1967).
At a sufficient distance from the epicenter where the curvature of  the
Rayleigh wave front can be neglected,  and with the proviso that
$\omega \gg \omega_a$,  acoustic waves are  emitted  at  the  angle
$\beta_R \approx arcsin(C_0/V_R) \approx 6^\circ$  to  the  vertical.

Rayleigh   waves   propagate   generally  in  the  form  of  a  train
consisting  of  several  oscillations  whose  typical   period   only
rarely   exceeds   several  tens  of  seconds.  Acoustic  waves  are
emitted  upward  in  the  form  of  the  same  train.  Because  of  a
strong absorption of the periodic wave, the only thing that is left
over in the case of the acoustic train  at  heights $z \geq$ 350~ km  is
the leading phase of compression.  It seems likely
that only in the case of  strong  earthquakes  (Alaskian
earthquake  of 1964),  even at large distances from the epicenter,
the disturbance (the leading  portion  of  the  acoustic  train)  that
remains  from  the  acoustic  train,  can  have  at  these altitudes a
duration of about  100  s,  and  quite  an  appreciable  ($u/C>$~0.1)
intensity (Orlov and Uralov, 1987) the nonlinear acoustics approximation.

Unfortunately, the parameters of Rayleigh waves in the neighborhood
of the subionospheric points appearing in
Table~3 are unknown to us. However, a most
pronounced bipolar character of the main signal
(Fig.~2b, Fig.~2e, Fig.~3b and Fig.~3e), its
intensity, and its long duration cast some doubt upon the fact that it is
the Rayleigh wave which is responsible for its origin. Such a
conclusion is consistent with the observed propagation direction of the
bipolar pulse, which makes a large angle with the vertical:
$\beta_{1,2} \approx 70$--$50^\circ \gg \beta_R^{"} \approx
15$--$18^\circ$.

Here it is taken into consideration that the acoustic ray originating from
a point on the ground at the angle of $\beta_R \approx 6^\circ$
now makes at the heights
$z=z_{eff}$ an angle $\beta_R^{"} > \beta_R$
because of the refraction effect  in  a  standard
atmospheric  model: $C_0/ sin~\beta_R = C / sin~\beta_R^{"}=const=V_R$,
$C(z = z_{eff}) \approx 0.9$--1~km/s.

The presence of strong winds at ionospheric heights can alter the value
of $\beta_R$. However, irrespective of the atmospheric model, the
phase velocity $V_h$ (Table~3) of the
horizontal trace of the acoustic disturbance generated by the Rayleigh
wave must coincide with $V_R$, and this is also not observed: $V_h
\ll V_R$.

\subsection{The epicentral emitter}
\label{EARTHQ-sect-5.2}

The detection of ionospheric disturbances, which are presumably
generated by a vertical displacement of the terrestrial surface directly
in the epicentral zone of an earthquake, using the GPS probing method,
is reported in (Calais and Minster, 1995).

Results of the present study lend support to the above conjecture.
However, the specific formation mechanism for the disturbance itself is
still unclear. An approach to solving this problem is contained in earlier
work, and involves substituting the epicentral emitter for a surface
velocity point source or an explosion. In particular, the substitution of
the earthquake zone for a point source turns out to be fruitful when
describing long-period internal gravity waves at a very long (thousands
of kilometers) distance from the epicenter (Row, 1967). The visual
resemblance of ionospheric disturbances at short (hundreds of
kilometers) distances from the earthquake epicenter to disturbances
from surface explosions is discussed in (Calais et al., 1998).

It should be noted that ionospheric disturbances generated by industrial
surface and underground nuclear explosions are also visually similar.
However, the generation mechanisms for disturbances are
fundamentally different in this case (Rudenko and Uralov, 1995). The
radiation source in underground nuclear tests is, as in the case of
earthquakes, the terrestrial surface disturbed by the explosion. The
intensity and spectral composition of the generated acoustic signal
reveal a strong (unlike the surface explosion) dependence on the zenith
angle, and are wholly determined by the form, size and characteristics
of the movement of the terrestrial surface in the epicentral zone of the
underground explosion.

In this Section,  we shall propose a model which,  we hope, will
help  to  understand   the   generation   mechanism   for   acoustic
disturbances --- the  subject of this paper.  Because of the
complexity of the problem formulated,  and for lack of  sufficient
data  on characteristics of the movement of the terrestrial surface
in epicentral  zones  of  earthquakes,  the  idealized  model  under
discussion   has   an   illustrative   character.   The  computational
scheme proposed  below  represents  a  simplified  variant  of  the
scheme   used   in Rudenko and Uralov,~(1995) to   calculate   ionospheric
disturbances generated by an underground confined nuclear explosion.

\subsection{The problem of radiation of the acoustic signal}
\label{EARTHQ-sect-5.3}

For the sake  of  simplicity,  we  consider  a  problem  having  an
axial  symmetry  about  the  vertical  axis $z$  passing  through the
earthquake epicenter $r=z=0$.  The epicentral emitter is a  set  of
plane  annular  velocity sources with the specified law of motion
along the vertical $U(r,t)$.  Since our interest is with the estimation
of  the  characteristics  of  an  acoustic  disturbance at a sufficient
distance from  the  emitter,  we  take  advantage  of  the  far-field
approximation    of    a    linear    problem   of   radiation.   In   the
approximation  of  linear acoustics $\omega \gg \omega_a$ and
with   no   absorption
present,  the gas velocity profile in the wave can be estimated by
the expression:

\begin{equation}
\label{EARTHQ-eq-10}
u(\tau, \beta)={\frac{A}{\pi R C_0}}\int_{0}^{L} \int_{-\infty}^{+\infty}
\frac{a(t^{'},r)rdrd{t^{'}}} {{\sqrt{y^2-{(\tau-t^{'})}^2}}\vert_{|\tau-t^{'}|\leq y}}
\end{equation}
where $\tau=t- \int_{0}^{l}{dl/C}$,~~$y=\frac{rsin\beta}{C_0}$.
Here $\beta$ is the zenith angle of departure of the acoustic ray from the
point $r=z=0$. $l$ is the group path length (the distance along the ray).
$a(t^{'},r)=\frac{dU(r,t)}{dt}$
is the vertical acceleration of the terrestrial surface.
$A=A(z)=\sqrt{\frac{{C_0}\rho_0}{C\rho}}$ is the acoustic factor.
$\rho_0,~\rho$ stand for the air density at the ground and at the  height
$z$, respectively, and L is a typical radius of the epicentral emitter.

In an  isothermal  atmosphere  where  the   ray   trajectories   are
straight  lines,  the quantity $R=(\sqrt{r^2+z^2}=l)$  is the radius
of a divergent (in the  far-field  approximation)  spherical  wave.
In this case the cross-section of the selected ray tube $S\propto R^2$.
In the real atmosphere
the value of $S$ is determined  from  geometrical  optics  equations,
and  using  the  expression~(\ref{EARTHQ-eq-10}) requires a further
complication of
the computational scheme.  Nevertheless,  to make estimates  we
shall use only the relation~(\ref{EARTHQ-eq-10}), and,  in doing so,
the value of $R$ will be corrected.

The case $\beta=0$ (and also $l=R=z$) is a special one:

\begin{equation}
\label{EARTHQ-eq-11}
u(\tau, \beta=0)=\frac{A}{RC_0} {\int_{0}^{L}a(\tau, r)rdr}
\end{equation}

Let us discuss the situation where all ring-type emitters `operate'
synchronously: $U(r,t)=U(t)$.  In this case the epicentral zone  is
emitting as a round piston 2$L$ in diameter.  Assume also that with
a  shock  of  the  earthquake,  the  vertical  displacement $\xi$,
the velocity $U=d\xi / dt$  and the acceleration $a=dU/dt$  are time
dependent, as shown in Fig.~4a.
For a rectangular velocity impulse,  i.e. in the limit
$\Delta t \gg \delta t \rightarrow 0$
(in this case $a\delta t \rightarrow \pm U_0 \delta(t^{'})$,
where $ \delta(t^{'})$  is a delta-function), from~(\ref{EARTHQ-eq-10})
we can obtain

\begin{equation}
\label{EARTHQ-eq-12}
u(\tau,\beta)=\frac{U_0 C_0 A}{\pi Rsin^2\beta}\lbrace{{\sqrt{y^2_L-\tau^2}}
\vert_{|\tau|\leq y_L}
-{\sqrt{y^2_L-{(\tau-\Delta t)}^2}}\vert_{|\tau-\Delta t|\leq y_L}}\rbrace
\end{equation}
where $y_L=\frac{Lsin\beta}{C_0}$;~~$U_0,~\Delta t$
are   the   amplitude  and  duration  of  the
rectangular   velocity    impulse.    In    this    case    the    vertical
displacement of the terrestrial surface after the earthquake shock
is $\xi_0=U_0 \Delta t$.  In the  strict  sense,  the
expression~(\ref{EARTHQ-eq-12}) holds  true  if  the
condition $y_L \gg \delta t$  is satisfied,  i.e.  with zenith angles
$\beta \gg \beta^{*} \approx arcsin\frac{C_0\delta t}{L}$.
Here $\delta t$ is the
operation duration of the terrestrial shock at the beginning of  the
movement and at the stop of the piston.
When $\beta \approx \beta^{*} \ll 1$ ,
the expression~(\ref{EARTHQ-eq-12}) and the
expression~(\ref{EARTHQ-eq-11}),
in view of $\delta t \neq 0$, yield approximately
identical  signals.  What  actually  happens  is that the validity of
the expression~(\ref{EARTHQ-eq-12})  will  break  down  still  earlier,
and  we  are justified in using it only for zenith angles
$\beta > \beta^{**}=arctan (\frac{L}{z}) >\beta^{*}$.

The curves in Fig.~4b
give  an  idea  of  the  relative  amplitude  and  form   of   acoustic
signals $u(\tau, \beta)$ in  the  far-field zone of radiation of the
piston $2L=60$~km in diameter.
The duration of the rectangular velocity impulse
of   the   piston   was   chosen   arbitrarily, $\Delta t=10$~s.
The  signals
correspond  to  the  expression~(\ref{EARTHQ-eq-12})  under  the  assumption  of  a
homogeneous, $A=1$, $C=C_{0} \approx 0.34$~km/s,  atmosphere.
The zenith angles are
$\beta=10^\circ, 20^\circ, 40^\circ, 60^\circ, 90^\circ$.
The spherical surface $R=const$ serves as a reference.
The abscissa axis indicates the time $\tau$ in seconds, and the axis of
ordinates indicates the gas velocity $u$.  In this case the  amplitude
of the most intense signal ($\beta=10^\circ$) is taken to be unity.

With increasing $\beta$, the amplitude of the signals $u_{max}$
decreases, and the duration $T$ increases. The leading and trailing
edges of the bipolar pulses in Fig.~2 and
Fig.~3 have an equal duration $\Delta t$. The
positive part of the bipolar pulse corresponds to the compression phase
of the acoustic wave, and the negative part refers to the rarefaction
phase. The area of the compression phase (in coordinates $u,t$) equals
the largest displacement $\chi_{+}$ of a unit volume of the atmosphere
in the direction of propagation (along the ray) of the wave. A total area
of a bipolar pulse is zero: $\chi_{+}= -\chi_{-}$. For acoustic signals,
described by the expression~(\ref{EARTHQ-eq-12}), the following
useful relations hold true:

\begin{equation}
\label{EARTHQ-eq-13}
u_{max}=U_0\frac{2AL}{\pi R
sin\beta}\sqrt{\eta-\eta^2};~~~T=2y_L+\Delta t
\end{equation}

\begin{equation}
\label{EARTHQ-eq-14}
\chi_{+}=\xi_0\frac{AL}{2\pi R sin\beta}\lbrace{\sqrt{1-\eta^2}+\frac{1}{\eta}
{arcsin~\eta}}\rbrace;~~~\eta=\frac{\Delta t}{2y_L}
\end{equation}

\subsection{The problem of acoustic signal propagation}
\label{EARTHQ-sect-5.4}

The wave vectors $\boldmath K_t$ of the bipolar pulses make with the
vertical  the  angles $\beta_1=70^\circ$  and $\beta_2=50^\circ$. With  the
adopted   values   of $z_{eff}=350$--400~km,   the   distances  of  the
corresponding   subionospheric   points    from    the    earthquake
epicenters  in Fig.~1a
are  approximately $r_1=800$~km and $r_2=600$~km.
The   rays,   constructed   in   the   approximation   of   linear
geometrical  acoustics  (LGA)  and  having  at  the heights
$z=z_{eff}$ ($C(z=z_{eff}) \approx 0.9$--1~km/s)
the   propagation   angles $\beta_1$ and $\beta_2$,
correspond to the zenith angle of departure
$\beta \approx 19$--$21^\circ$ and $\beta \approx 15$--$17^\circ$
at the level $z=0$.

The fact that these values of $\beta$ satisfy the inequality
$\beta \leq \beta^{*} \approx 25$--$30^\circ$,
is in reasonably good agreement with the familiar picture of
rays from  a  ground-level  point  source  ---  the  rays  with
$\beta \geq \beta^{*}$  are
captured by the atmospheric waveguide $z \leq z^{*}=120$~km,  and only
the rays emitted upward inside  the  solid  angle
$\Omega^{*} \approx1$~sterad  can penetrate   to   the
heights $z~>~z^{*}$.
In   standard   models  of  the
atmosphere,  however,  to  the  values  of   the   angles
$\beta_{1,2}(z=z_{eff})$ there
correspond  the  locations  of subionospheric points lying several
hundreds kilometers nearer to the  epicenter  compared  with  the
experimental   values   of $r_1 \approx 800$~km  and  $r_2 \approx 600$~km.
This inconsistency can be due to two reasons.

One reason is that the value of $z_{eff} \approx 350$--400~km, which we are
using,  is too low. This is supported by the detection of velocities
$V_t \approx 1.2$~km/s of traveling disturbances
(Table~3)  which  exceed
markedly  the  sound  velocity  $C\approx 0.9$--1~km/s  at the heights of
$\approx 350$--400~km.  However, it seems likely that such a discrepancy
may  be  disregarded,  in  view  of  the errors of the measurement
technique used (the probability  of  an  additional  heating  of  the
upper  atmosphere  prior  to  the  earthquake  cannot be ruled out,
however).

The increase of the actual value of $z_{eff}$ can also be associated
with a strong dependence (see~(\ref{EARTHQ-eq-13}),
(\ref{EARTHQ-eq-14}) and Fig.~4b) of the power
of the emitted signal on the zenith angle of departure of the ray from
the earthquake epicenter. Verifying this factor requires a more detailed
analysis, based on particular data on vertical movements of the
terrestrial surface in the epicentral zone. Such data are unavailable to
us. Below is a discussion only of the second reason for the
above-mentioned inconsistency as a highly probable one.

The second reason may be associated with  the  violation  of  the
validity   conditions  of  the  LGA  approximation  at  a  sufficient
distance  from  the  source  of  the  acoustic   disturbances   under
discussion.   Indeed,   the   utilization   of   this   approximation  is
justified until the  parameters  of  the  medium  and  of  the  wave
itself change substantially on the size of the first Fresnel zone
$d_F \approx \sqrt{\lambda l}$  that  determines  the  physical
(transverse)  size  of  the   ray.
Typical  wavelengths  of  the  bipolar  pulses  under discussion at
the heights $z=z_{eff}$ are large: $\lambda \approx 300$--200~km.
Distance $l$  along  the  expected  ray  is  of  order  900--700~km.
Then $d_F \approx 520$--370~km,   which   exceeds   substantially
the   scales   of
variation  of atmospheric parameters.  The value of $d_F$ is actually
somewhat smaller,  because  the  typical  scale  of  a  disturbance
decreases  as  it  approaches  the  source.  A violation of the LGA
approximation   at   the   above-mentioned   distances   from    the
epicentral  source  occurs  also for model signals
$\beta \geq 20^\circ$,  shown in Fig.~4b.

The increasing violation  of  the  applicability  conditions  of  the
LGA  approximation  with an increase of $l$ implies the transport
of the wave  energy  not  strictly  along  the  calculated  rays  but
along the lines with a smaller curvature. With a mere estimate of
the dilution factor $R$ in  the
expressions~(\ref{EARTHQ-eq-12}),~(\ref{EARTHQ-eq-13})
and~(\ref{EARTHQ-eq-14}), in
mind,  these  lines  will be assumed to be straight when $z>z^{*}$ and
to originate  from  an  imaginary  source  lying  at  the  height
$z^{*}\approx 120$~km above the earthquake epicenter.
Such a situation is also
clearly manifested  in  the  LGA  approximation.  At  the  heights
$z>z^{*}$,  at $t \approx 600$~s,  for example,
the surface of the wave front
from the ground-level impulsive source resembles the surface  of
a hemisphere centered on the point $r \approx 0$, $z \approx z^{*}$.

Ultimately the  energy  that  arrives  from  below inside the solid
angle $\Omega^{*} \approx 1$ is scattered into a solid angle
$\Omega \approx 2\pi$. From  the  condition  of
conservation of wave energy, it is possible to find
$R_{1,2}=\sqrt{\frac{\Omega}{\Omega^{*}}} \sqrt{{(\Delta
z)}^2+r^2_{1,2}}$, when $\Delta z=z_{eff}-z^{*}$.
When using
the expressions~(\ref{EARTHQ-eq-13}),~(\ref{EARTHQ-eq-14}) in  the
subsequent  discussion,  we
will  take  the  quantity  $R_1$  rather  than  $R$,  and the value of the
zenith  angle  of  departure  will  be  taken  to  be $\beta=20^\circ$.
These parameters   approximately   correspond   to   the  generation  and
propagation conditions of the  signal  from  the  first  earthquake.
The  typical  size  of  the  epicentral  zone of this earthquake was
about $2L=60$~km (according to the USGS data: www.neic.cr.usgs.gov).
This   same   value   of  $L$   was   used   in
calculating the signals shown in Fig.~4b.

As is evident even from Fig.~4b (thick line),
the duration  of  the
signal,  having  a  zenith angle of departure $\beta=20^\circ$,
is about 70~s.
When  the  signal  propagates  in  the   approximation   of   linear
acoustics  and  with  no  absorption,  its form and duration remain
unchanging, and only its amplitude changes.

In actual   conditions   the   combined   effect   of   the   nonlinear
attenuation  and  linear  absorption factors leads to a stretching of
the bipolar pulse,  and to a change of its form (we do not  discuss
the  dispersion  factor).  In  this case the effect of the nonlinearly
factor occurs in such a manner that the integral value of
$\chi_{+}$,~(\ref{EARTHQ-eq-14}),
calculated  as  an  approximation  of  linear acoustics,  remains as
such  in   the   approximation   of   nonlinear   acoustic   as   well.
Moreover,  taking  into  account  the finite width $\Delta T_{sh}$
of the shock
front  does  not  change  this  situation   until $\Delta T_{sh}<T/4$.
The   linear absorption factor,  in turn,  somewhat reduces the
true value of $\chi_{+}$
because   of   the   mutual   diffusion   of   the   compression    and
rarefaction  phases.  Nevertheless,  for  a hypothetical estimation
of the earthquake parameters,  we shall use the assumption about
the conservation of the value of $\chi_{+}$.

As is intimated by Table~3, the mean   value   of  the $TEC$
disturbance  amplitude  after  the  first  earthquake  is
$A_I \approx 0.14$ $TECU$  at the equilibrium value of
$I \approx 5$ $TECU$.  Assuming that on
the order of magnitude $A_I/I \approx u/C$ for a maximum gas velocity  in
the  wave,  we have an estimate of $u^{exp} \approx 30$~m/s.
For the sinusoidal
form of the bipolar pulse with a duration $T^{exp} \approx 350$~s,
we  find  the
experimental  value  of  gas  displacement  along  the direction of wave
propagation: ${\chi_{+}}^{exp}=u^{exp}T^{exp}/\pi \approx 3$~km.
Using  the  relation ${\chi_{+}}^{exp}\approx \chi_{+}$ it  is
possible   to
estimate  the  vertical displacement $\xi_0$ of the terrestrial surface in
the epicentral zone of the earthquake.  For this purpose,  it seems
reasonable  to  introduce  the  assumption  about  a short velocity
impulse which models the main earthquake shock, $\eta \ll 1$, although
the  numerical value of $\chi_{+}$ is virtually independent of the value of
$\eta$~(\ref{EARTHQ-eq-14}). In view of the above considerations,
we then obtain:

\begin{equation}
\label{EARTHQ-eq-15}
\xi_0 \approx {\chi_{+}}^{exp}\pi R_{1}sin\beta/(AL).
\end{equation}

The uncertainty in the determination of $\xi_0$ is caused both  by  the
uncertainty  of  the  true  values  of the quantities involved in this
relation and by the limitations of the  acoustic  signal  generation
model   itself.   In  this  case,  of  the  greatest  importance  is  the
dependence  of $\xi_0$ on  the  value  of  the  acoustic  parameter $A$,
containing  the  atmospheric  density $\rho$ at  the  effective  height
$z=z_{eff}$ which we have introduced  artificially.  The  employment
of   the   MSISE90   atmospheric   model (Hedin, 1991), calculated  for  the
location and time of the first earthquake,  gives the  values  of
$\xi_0 \approx 60, 40$ and 25~cm,  with the values of
$z_{eff}=350, 400$ and 450~km,  respectively.
In all cases it was assumed that
$R_{1} \approx 850 \sqrt {2\pi}$~km, $\beta=20^\circ$, $L=30$~km.
The possibility of a vertical displacement of the
terrestrial  surface  in  the  epicentral  zone  by   several   tens   of
centimeters seems real.  In particular, (Calais and Minster, 1995) give
the value of $\xi_0 \approx 40$~cm at the epicenter of the Mw=6.7
Northridge earthquake (California,  1994).

In view of the demonstration character of the above calculation, we
have intentionally excluded from consideration the effects associated
with the inclination of the magnetic field lines, and with the possible
presence of strong winds at ionospheric heights. The presence of a
magnetic field modifies the picture of transfer of movements from the
neutral gas to the electron component of the ionosphere. Since the
magnetic field is not entrained by the neutral gas, the field lines can be
considered fixed. In this case the acceptable approximation would be
the one, in which the electron component travels only along magnetic
field lines with the velocity $u\cdot cos\psi$, where $\psi$ is the angle
between the magnetic field vector and the velocity vector of the neutral
gas. Therefore, the quantity $u\cdot cos\psi$ must be involved in lieu
of the quantity $u$ in the expression $A_I/I \approx u/C$ that was used
above.

Let us estimate the value of $\psi$ for Turkey earthquakes.
In the examples under discussion (Table 3),
the horizontal projection $\boldmath K$ of the full wave
vector $\boldmath K_t$ is virtually collinear to the horizontal
component of the magnetic field. The wave vectors $\boldmath K_t$,
in turn, make with the vertical the angles from $\beta_1\approx
70^\circ$ to $\beta_2\approx 50^\circ$.

Since the magnetic dip in the middle of Turkey is about $60^\circ$ (the
angle is measured from the horizontal plane), $\psi\cong
20^\circ$--$40^\circ$, and the value of $cos\psi\approx 0.94$--$0.77$
differs little from 1. It should be remembered, however, that taking into
account this factor can be very important in the analysis of the
complete picture of TEC disturbances above the earthquake or
explosion source (see, for example, Calais et al., 1998).

The presence of the zonal and meridional winds at ionospheric heights
leads to a displacement and deformation of the wave front, and hence
give rise to a dependence of the acoustic wave intensity on the
propagation direction. The decisive role in this case is played by the
wind velocity gradient. This factor can be taken into account within the
framework of the ray theory. However, a corresponding model
calculation would be worthwhile in the analysis of experimental data
obtained for a set of subionospheric points surrounding the acoustic
wave source. In the present situation, however, where the number of
subionospheric points used in the analysis is too small, and the
uncertainty of the parameters of the acoustic emitter itself is too large,
the solution of such an unwieldy problem would be an overrun of the
accuracy which is pursued by the above computational scheme.

As follows from the expressions~(\ref{EARTHQ-eq-13}),
maximum values of displacements and  of  the
velocity  of the  neutral  atmospheric  species  are  attained
directly above the earthquake epicenter.
The signal duration is  minimal,  and  does
not  seem to exceed a few tens of seconds at ionospheric heights.
Since   the    wave    vector    of    the    disturbance    is    directed
predominantly    upward,    the    method    of    oblique-incidence
ionospheric sounding in this  case  is  a  technique  of  choice  for
determining the waveform.

\section{Conclusion}
\label{EARTHQ-sect-6}

In this paper we have investigated the form and dynamics of
shock-acoustic waves generated during earthquakes. We have
developed a method of determining the SAW parameters using GPS
arrays whose elements can be chosen out of a large set of the global
network GPS stations. Unlike existing radio techniques, the proposed
method estimates the SAW parameters without a priori information
about the location and time of the earthquake. The implementation of
the method is illustrated by an analysis of ionospheric effects of the
earthquakes in Turkey (August 17, and November 12, 1999), in
Southern Sumatera (June 4, 2000), and off the coast of Central America
(January 13, 2001).

It was found that, in spite of the difference of the earthquake
characteristics, the local time, the season, and the level of
geomagnetic disturbance, for four earthquakes the time period of
the ionospheric response is 180-390 s, and the amplitude exceeds
by a factor of two as a minimum the standard deviation of
background fluctuations in total electron content in this range
of periods under quiet and moderate geomagnetic conditions.

As has  been  pointed  out in the Introduction,  some investigators
report  markedly   differing   values   of   the   SAW   propagation
velocity   ---   by   as  much  as  several  thousands  m/s,  which  is
beyond the values of the sound velocity at the SAW propagation
heights in the atmosphere. The method proposed in this paper opens up a
possibility of determining the angular characteristics of the wave vector
$\boldmath K_t$ and, accordingly, of estimating $V_t$. According to our data
(Table~3), the elevation of the SAW wave vector varied within
20--$44^\circ$ , and the phase velocity of the
SAW varied from 1100 to 1300~m/s. We determine the phase velocity
of the equal TEC line at the height of the ionospheric F-region
maximum which makes the main contribution to variations of the TEC
between the receiver and the GPS satellite, and corresponds to the
region of maximum sensitivity of the method. Since $V_t$ approaches
the sound velocity at these heights (Li et al., 1994) this makes it
possible to identify the sound origin of the TEC disturbance. The SAW
source location, calculated without taking into account the refraction
corrections, approximately corresponds to the earthquake epicenter.

\section{Acknowledgements}
\label{EARTHQ-sect-7}

We are indebted to G.~M.~Kuznetsova and A.~V.~Tashchilin for their
calculation of the atmospheric parameters at the time of the
earthquakes, as well as to E.~A.~Ponomarev, V.~V.~Evstafiev,
P.~M.~Nagorsky, N.~N.~Klimov, and A.~D.~Kalikhman for their
interest in this study, many pieces of useful advice, and active
participation in discussions. Thanks are also due to
V.~G.~Mikhalkosky for his assistance in preparing the English version
of the manuscript. Finally, the authors wish to thank the referees for
valuable suggestions which greatly improved the presentation of this
paper. This work was done with support from both the Russian
foundation for Basic Research (grant 99-05-64753) and RFBR grant of
leading scientific schools of the Russian Federation No. 00-15-98509.

\end{document}